\documentclass[twoside,12pt]{article}
\usepackage{macros2e}

\usepackage{newcomnd,din_a4,epsf,graphicx,pictures,amssymb}
\newlength{\pcm}
\newlength{\pmm}
\newcommand{\contract}{\;\raisebox{0.2ex}{\scalebox{0.6}{\rotatebox{90}{$\blacklozenge$}}}\;}
\newcommand{\suppress}[1]{}
\newcommand{\suppressmargin}[1]{}

\newcommand{\rD}{\mathrm{D}}

\usepackage{times}

\newcommand{\glue}{\hspace*{-0.17ex}}

\begin{document}
\begin{titlepage}
\noindent
\renewcommand{\thefootnote}{\fnsymbol{footnote}}
\parbox{1.85cm}{\epsfxsize=1.85cm \epsfbox{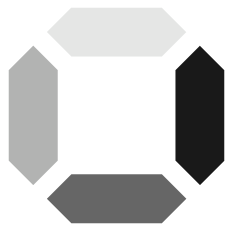}}\hfill%
\begin{minipage}{10cm}
\rightline{Uni GH Essen preprint}%
\rightline{chao-dyn/9911005v3}%
\rightline{July 7, 2000}%
\end{minipage}
\vfill 
\centerline{\sffamily\bfseries\Large The passive polymer problem}
\vfill
\centerline{\bf\large Kay J\"org Wiese%
\footnote{Email: wiese@next23.theo-phys.uni-essen.de}}
\smallskip
\centerline{\small Fachbereich Physik, Universit\"at GH Essen,  45117 Essen,
Germany}
\smallskip\smallskip

\vfill
\vspace{-5mm}
\begin{abstract}
In this article, we introduce a generalization of the diffusive
motion of point-particles in a turbulent convective flow with 
given correlations to a polymer or membrane. In analogy
to the {\em passive scalar problem} we call this 
the {\em passive polymer} or {\em membrane} problem.
We shall focus on the expansion about the 
marginal limit of velocity-velocity correlations
which are uncorrelated in time
and grow with the distance $x$ as $|x|^\E$, and $\E$ small.  
This relation gets modified in the case of polymers and membranes (the 
marginal advecting flow has correlations which are shorter ranged.)
The construction is done in three steps: First, we reconsider the
treatment of the passive scalar problem using the most convenient treatment 
via field theory and renormalization group. 
We explicitly show why IR-divergences and thus the system-size
appear in physical observables,
which is rather unusual in the context of ordinary field-theories, like 
the $\phi^4$-model. We also discuss, why the renormalization group 
can nevertheless be used to sum these divergences and leads to 
anomalous scaling of $2n$-point correlation functions as e.g.
$S^{2n}(x):=\langle \left[ \Theta(x,t)-\Theta(0,t) \right]^{2n}  \rangle$. 
In a second step, we reformulate the problem in terms of a 
Langevin equation. This is interesting in its own, since it allows
for a distinction between single-particle and multi-particle 
contributions, which is not obvious in the Focker-Planck treatment. 
It also gives an efficient algorithm to determine $S^{2n}$ numerically,
by measuring the diffusion of particles in a random velocity 
field. 
In a third and final step, we generalize the Langevin treatment
of a particle to polymers and membranes, or more generally to 
an elastic object of inner dimension $D$ with $0\le D \le 2$.
These objects can intersect each other. 
We also analyze what happens when self-intersections are no longer
allowed. 
 


\end{abstract}
\vspace{-5mm}
\vfill

\centerline{\em Accepted for publication in J.\ Stat.\ Phys.} 

\vfill

\end{titlepage}
\renewcommand{\thefootnote}{\fnsymbol{footnote}}
{
\setcounter{tocdepth}{4}
\tableofcontents} \newpage
\setcounter{page}{3}
%
\newcommand{\sureinput}[2]{\protect\markboth{K.\,J.~Wiese, The passive polymer problem}{#2}\input{#1}}
\newcommand{\maybeinput}[2]{\protect\markboth{K.\,J.~Wiese, The passive polymer problem}{#2}\input{#1}}

\renewcommand{\thefootnote}{\arabic{footnote}}
\setcounter{footnote}{0}
\protect\markboth{K.\,J.~Wiese, The passive polymer problem}{Introduction and outline}\section{Introduction and Outline}
\begin{figure}[t]
\centerline{\parbox{0cm}{\hspace*{3cm}\epsfxsize=4cm\epsfbox{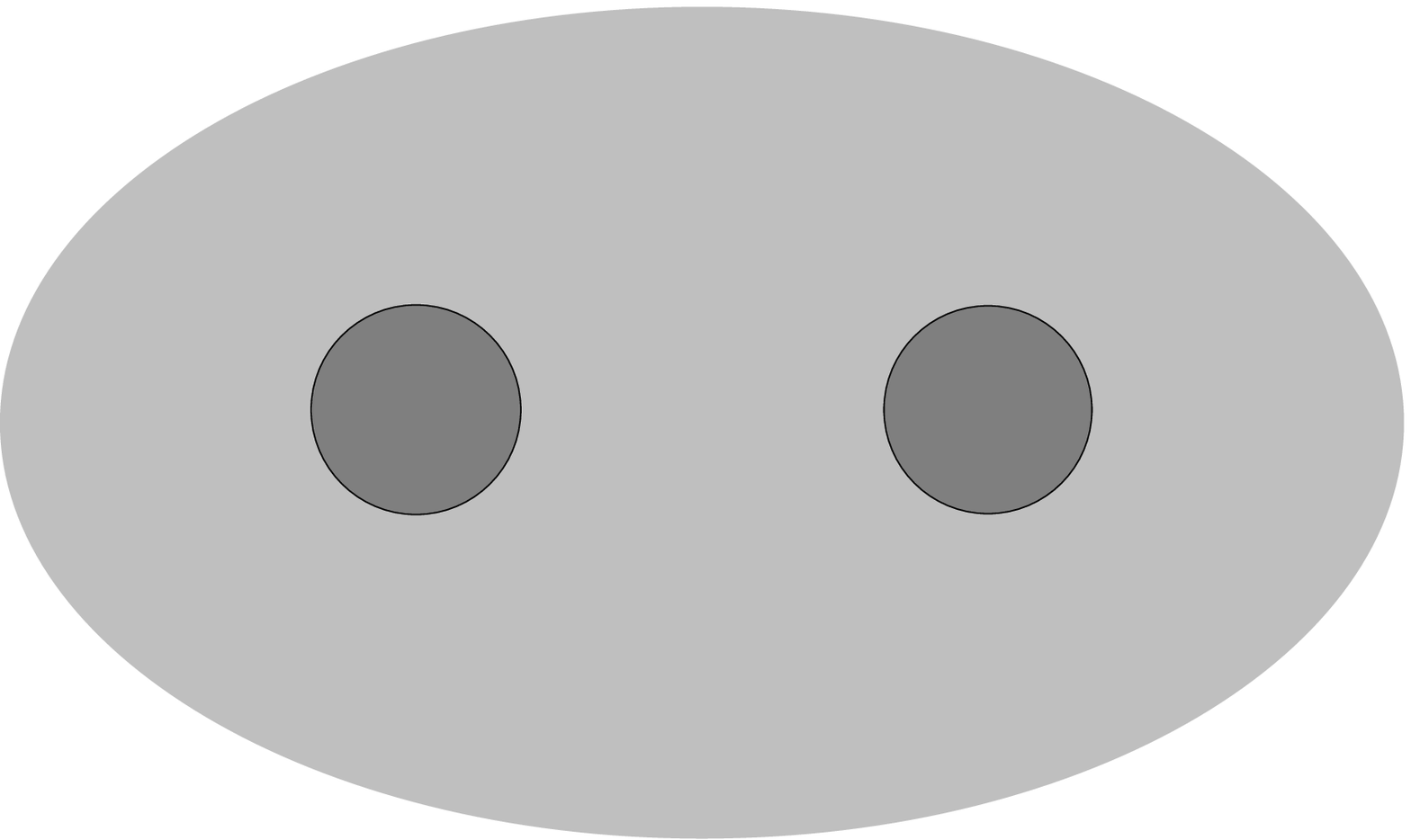}}%
\parbox{10cm}{\epsfxsize=10cm\epsfbox{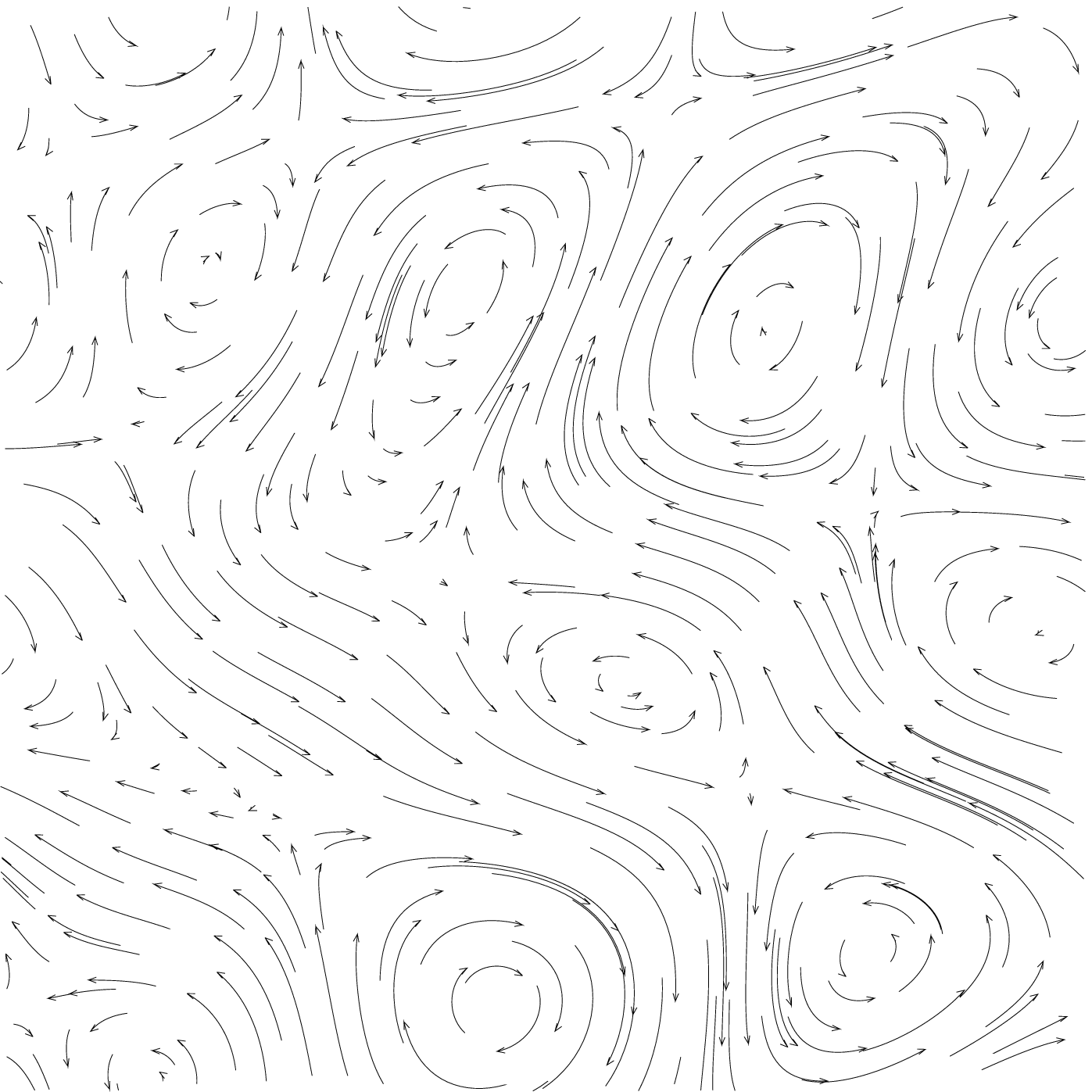}}}
\caption{Symbolic picture of a turbulent flow. 
Particles, or equivalently heat is injected in 
a finite range of size $L\sim 1/M$ (dark grey areas), 
whereas the turbulent flow grows
up to scale $l\sim 1/m$, which finally shall be taken to 
infinity. This is possible, if the total number of particles, or the 
total heat, injected into the system is conserved. In that case, 
$L$ and not $l$ sets the largest scale in the problem, and multi-point
correlation functions with an anomalous $L$-dependence
will be observable in a domain of size $L$, 
here symbolically
 shaded in light grey. As will be shown below, they  have anomalous 
corrections depending on $L$. }
\label{i:figure turb flow}
\end{figure}
For now more than 5 decades, turbulence has resisted a satisfying theoretical
treatment. The principle question asked since Kolmogorov's pioneering
work \cite{Kolmogorov1941} in 1941 is whether there are corrections to the 
simple scaling behavior predicted in 
\cite{Kolmogorov1941} for higher correlation functions \cite{Oboukhov1962,%
Kolmogorov1962}. 
The most natural tool to answer this question is the renormalization 
group. However, all attempts to go beyond Kolmogorov's analysis 
have essentially failed so far. To better pin down
the  problem, simpler toy models have been proposed.
The probably most prominent such model is the {\em passive scalar} model, 
introduced by Obukhov \cite{Obukhov1949} and  Kraichnan 
\cite{Kraichnan1968}. 
This model describes the diffusion of a point-particle
in a turbulent flow with given correlations. For simplicity these 
correlations are taken to be Gaussian. Nevertheless, the model 
is far from beeing simple, and shows multi-scaling, i.e.\ higher correlation
functions of the particle density scale independly of the 
second moment, characterized by new critical exponents. 
More explicitly, particles, or equivalently heat is injected in 
a finite range of size $L\sim 1/M$, whereas the turbulent flow grows
up to a bound of $l\sim 1/m$, which finally shall be taken to 
infinity. This is possible, if the total number of particles, or the 
total heat, injected into the system is conserved. In that case, 
$L$ and not $l$ sets the largest scale in the problem, as visualized
in figure \ref{i:figure turb flow}.

In this article, we introduce the generalization from point particles
to higher dimensional elastic objects, as e.g.\ polymers and membranes. 
In analogy
to the {\em passive scalar} problem we call this 
the {\em passive polymer} or {\em passive membrane} problem.
%
 
We start by considering
 the passive polymer problem.
Much has been learned during the last years about
higher correlation functions due to a common 
effort of mathematicians and physicists 
\cite{ChenKraichnan1989,GawedzkiKupiainen1995,%
GawedzkiKupiainen1995b,ChertkovFalkovichKolokolovLebedev1995,%
KraichnanYakhotChen1995,%
ChertkovFalkovich1996,BernardGawedzkiKupiainen1996,%
ShraimanSiggia1996,%
ChertkovFalkovichLebedev1996,%
BalkovskyLebedev1998,AdzhemyanAntonovVasilev1998,%
LvovProcacciaFairhall1994,FairhallGatLvovProcaccia1996,%
GatLvovProcaccia1997,GatLvovPodivilovProcaccia1997,%
FairhallGalantiLvovProcaccia1997,BernardGawedzkiKupiainen1998,%
ChenKraichnan1998,GawedzkiVergassola1998,%
GatZeitak1998,GatProcacciaZeitak1998,%
FrischMazzinoNoullezVergassola1999}. 
Whereas the first to calculate the 4-point function
by considering the 0-modes of the steady state
are \cite{GawedzkiKupiainen1995}, the calculatory most
convenient scheme, based on the perturbative renormalization group,
was introduced in \cite{AdzhemyanAntonovVasilev1998}. 
Contrary to the sometimes heard claim,  
the renormalization group is able to handle large eddy motion. 
The  expansion is performed about the 
marginal limit of velocity-velocity correlations
which are uncorrelated in time
and grow with the distance $x$ as $|x|^\E$, and $\E$ small,  
a relation which gets modified for polymers and membranes (the 
marginal advecting flow has correlations which are shorter ranged.)

The generalization to polymers and 
membranes  is then performed  in three steps: First, we reconsider the
treatment of the passive scalar problem using the most convenient treatment 
via field theory and renormalization group. 
We explicitly show why IR-divergences and thus the system-size
appear in physical observables,
which is rather unusual in the context of ordinary field-theories, like 
the $\phi^4$-model. We also discuss, why the renormalization group 
can nevertheless be used to sum these divergences and leads to 
anomalous scaling of $n$-point correlation functions as e.g.
$S^{2n}(x):=\langle \left[ \Theta(x,t)-\Theta(0,t) \right]^{2n}  \rangle$.
To do so, we determine the full scaling dimension of the composite
operators 
${\cal S}^{(n,m)}:=  [(\nabla \Theta)^2]^n\, [z\nabla \Theta]^{2m}$, 
with $|z|=1$.
In a second step, we reformulate the problem in terms of a 
Langevin equation. This is interesting in its own, since it allows
for a distinction between single-particle and multi-particle 
contributions, which is not obvious in the Focker-Planck treatment. 
It also gives an efficient algorithm to determine $S^{2n}$ numerically,
by measuring the diffusion of particles in a random velocity 
field. 
In a third and final step, we generalize the Langevin treatment
of a particle to polymers and membranes, or more generally to 
an elastic object of inner dimension $D$ with $0\le D \le 2$.
Our analysis will show that the interesting range for $\E$ is 
$-\frac{2D}{2-D}<\E<0$. For smaller $\E$, the advecting flow
is irrelevant. For larger $\E$, the polymer or membrane is 
overstretched. This is the range, where already the particle, 
i.e.\ the center of mass of the polymer, shows anomalous diffusion. 
We also generalize these considerations to the case of self-avoiding
polymers and membranes. 



\protect\markboth{K.\,J.~Wiese, The passive polymer problem}{The passive scalar}\section{The passive scalar}
\subsection{Model}
\label{s:Model}

The advection of a passive scalar field $\Theta(x,t)$ with $x\in \R^d$
the spatial coordinate and $t$ the time, is described by 
the Focker-Planck type equation
\cite{Obukhov1949,%
Kraichnan1968}
\be \label{s:FPE}
\left[ \p_t +v(x,t) \nabla \right] \Theta(x,t) =\nu_0 \Delta \Theta(x,t)+f(x,t)
\ .
\ee
The correlations of the advecting turbulent
 velocity field $v(x,t)$ are supposed to be Gaussian with zero mean
 and correlations which grow with the distance $r$ as $r^\E$
\be \label{s:vvcorrel}
\left< v^i(x,t) v^j(x',t') \right> = \rD_v^{ij}(x-x',t-t') =\rD_0 \delta(t-t') \int\!
\frac{\rmd^d k}{(2\pi)^d}\, P^{ij}(k) \,\frac{\rme^{ik(x-x')}}{(k^2+m^2)^{\frac{d+\E}2}}\ ,\quad
\ee
where 
\be
P^{ij}(k):=\delta^{ij}-\frac{k^ik^j}{k^2}
\label{s:trans Project}
\ee
is the transversal projector and $m$ some IR-regulator.  The dimension
of the coupling $u_0:=\rD_0/\nu_0$ in units of $m$ is 
\be
\E= \left[u_0 \right]_m \ .
\ee
We will see later that $\E$ serves as a regulator. Eventually
one is interested in the physically relevant case of $d=3$ and 
$\E=2/3$ (Kolmogorov-scaling) or corrections thereto \cite{Landau6}.

$f$ is a Gaussian scalar noise with zero mean and 
correlator
\be \label{s:f-correl}
\left< f(x,t) f(x',t') \right> = 
\delta(t-t') G_f(Mx, Mx') \equiv \delta(t-t') G_f^M(x-x')\ .
\ee
Often people use $G_f(Mx, Mx')=G_f(M|x-x'|)$, and 
its Fourier-transform $\tilde G_f(k/M)$. However we will use
the more general case for clearness of derivation. 
$G_f$ is the  source of correlation-functions of  $\Theta$, 
which otherwise would vanish. Physically, it may be viewed 
as source and sink of tracer-particles, or heat. We will see below
when explicitly calculating expectation values that $G_f$ 
sets the largest scale $L\equiv1/M$ appearing in physical observables. Therefore, we demand that $G_f(s,s')$ rapidly decays to zero for 
$s$ or $s'$ larger than 1.

The analysis of \Eq{s:FPE} is most easily done by using a dynamic action 
\cite{MSR,Janssen1992}
\be \label{s:dynamic action with v}
J[\Theta,\tilde \Theta,f,v] = 
\int\limits_{x,t} \tilde \Theta(x,t) \left[
\p_t  \Theta(x,t) -\nu_0 \Delta   \Theta(x,t) - v(x,t)   \nabla \Theta(x,t) 
-f(x,t)\right] 
\ .
\ee
Expectation values are obtained by integrating
 $\rme^{-J[\Theta,\tilde \Theta,f,v]}$ over $\Theta,\tilde \Theta$ 
 and averaging over $f$ and $v$.
Since  $v$ and $f$ are Gaussian, their average can be taken, leading
to
\be \label{s:dynamic action}
J[\Theta,\tilde \Theta] = \int_{x,t} 
\tilde \Theta(x,t) (\p_t-\nu_0\Delta) \Theta(x,t) 
- \int_{x,y,t}\left[  \frac{\rD_0}2 \DA + \half \PD\right]\ ,
\ee
where
\bea
\rD_0 \DA &=& \tilde \Theta(x,t) \nabla_i \Theta(x,t)\, \tilde \Theta(y,t) \nabla_j \Theta(y,t) \rD_v^{ij}(x-y)\\
&=& \Theta(x,t) \nabla_i \tilde \Theta(x,t)\, \Theta(y,t) \nabla_j \tilde \Theta(y,t) \rD_v^{ij}(x-y) = \rD_0 \DB\nn\\
\PD &=& \tilde \Theta(x,t)\tilde \Theta(y,t) G_f(x,y) \nn
\ .
\eea
Note that due to the transversal
projector in the turbulent interaction, the partial integration 
from $\int_x \tilde \Theta(x,t) \nabla_i \Theta(x,t) D^{ij}(x-y)$
to $-\int_x \nabla_i \tilde \Theta(x,t)  \Theta(x,t) D^{ij}(x-y)$
is possible since $\nabla_i \rD_v^{ij}(x-y) =0$. 

The free response and correlation functions read
\bea
\EXP{ \tilde \Theta(k,\omega) \Theta(k',\omega') }0 &=&
(2\pi)^{d+1}\delta(\omega+\omega')  \delta^d(k+k')
R(k,\omega)\ , \quad R(k,\omega) = \frac{1}{i\omega+\nu_0 k^2}\nn\\
\EXP{\Theta(k,\omega) \Theta(k',\omega') } 0&=&
(2\pi)^{d+1}\delta(\omega+\omega')  \delta^d(k+k')
C(k,\omega)\ , \quad C(k,\omega) = \frac{\tilde G_f(k)}{\omega^2+(\nu_0 k^2)^2}
\ , \nn\\
\eea
where in the last formula $G_f(x,x')$ was supposed to be of the form
$G_f(x-x')$.
Most convenient is  a mixed
time and $k$-dependent representation
\be \label{s:R mixed}
\EXP{ \tilde \Theta(k,t) \Theta(k',t') }0 = (2\pi)^{d} \delta^d(k+k')
R(k,t'-t)\ , \quad R(k,t) = \Theta(t) \rme^{-\nu_0 k^2 t}
\ .
\ee
This also  yields the response-function in position space
\be \label{s:R(x,t)}
R(x,t) = \Theta(t)\, (4\pi  \nu_0 t)^{-\frac d2}\, \rme^{-\frac{x^2}{4\nu_0 t}}
\ .
\ee

\subsection{Perturbative corrections, renormalization of the dynamic action $J$}
\label{s:Perturbative corrections...}
We now study the renormalization of the model, i.e.\ we want 
to eliminate all UV-divergent terms at $\E=0$. It is important to
notice that such divergences only
come from the insertion of the turbulence-interaction 
$\frac{\rD_0}2 \DA$, but {\em not} from the insertion of the source
of tracer-particles $\half \PD$.
To first order in $\rD_0$, the only contribution is
\be
\rme^{-J} ~\longrightarrow~ \frac{\rD_0}2 \DA ~\longrightarrow~ -\rD_0\PA  \ .
\ee
The diagram is without the external legs in the most convenient
 mixed $t$ and $k$  
representation of \Eq{s:R mixed}
\be \label{s:nu ren}
\PB = \int\limits_{-\infty}^\infty\rmd t \int \frac {\rmd^d k}{(2\pi)^d}\,
\Theta(t) \rme^{-\nu_0 k^2 t }  \frac{\delta_\eta(t)}{(k^2+m^2)^{\frac{d+\E}2}}
\left(1-\frac1d\right)\ .
\ee
In order to clarify the role of the factor $\delta(t)$ in $\DA$, 
we have to recall that this is an approximation for a sharply 
peaked but nevertheless smooth function around $t=0$. This is the reason 
why in \Eq{s:nu ren}, we have replaced the $\delta$-distribution 
by a smoothened one $\delta_\eta(t)$, which in the limit of $\eta\to0$
will reproduce $\delta(t)$. Integrating 
$\int \rmd t \, \theta(t) \rme^{-k^2\nu_0 t} \delta_\eta(t)$ and then 
taking the limit of $\eta\to 0$ thus simply yields a factor of $\half$. 
\Eq{s:nu ren} becomes
\be
 \PB = \half \int \frac {\rmd^d k}{(2\pi)^d}\,
  \frac1{(k^2+m^2)^{\frac{d+\E}2}}
\left(1-\frac1d\right) = \half\left(1-\frac1d\right) C_d  \frac{m^{-\E}}\E\ ,
\ee
where $C_d$ is defined as 
\bea \label{s:Cd def}
C_d &:=& \E \int \frac {\rmd^d k}{(2\pi)^d}\,
  \frac1{(k^2+1)^{\frac{d+\E}2}} \nn\\
  &=& \frac{2\Gamma(1+\frac\E2)}{\Gamma(\frac{d+\E}2) (4\pi)^{d/2}}
\ .
\eea
This leads at leading order 
to a renormalization of $\nu$ (denoting with subscript 0
bare quantities)
\bea \label{s:Znu}
\nu_0 &=& \nu Z_\nu  \nn\\
Z_\nu &=& 1- \frac{u}\E \left(1-\frac1d\right) \frac{ C_d}2 \ ,
\eea
where we have introduced a coupling $u_0$ and its renormalized
counterpart $u$ through
\bea \label{s:u}
D_0 &=&Z_D D  \\
u_0 &=& \frac {\rD_0}{\nu_0} = u m^{\E} \frac{Z_\ind D}{Z_\nu}
\ .
\eea
We now claim that \Eqs{s:Znu} and \eq{s:u} are all renormalizations
 needed, and that even to all orders in perturbation theory. 
Let us first focus on the renormalization of $\rD$. There will appear diagrams
like
\be
{{\stackrel{q}\longrightarrow}\atop{\stackrel{q+l}\longleftarrow}}
 \bareDB\glue\bareDA
 {{\stackrel{p}\longrightarrow}\atop{\stackrel{p+l}\longleftarrow}}
 ~\equiv~
 {{\stackrel{q}\longrightarrow}\atop{\stackrel{q+l}\longleftarrow}}
 \bareDA\glue\bareDB
 {{\stackrel{p}\longrightarrow}\atop{\stackrel{p+l}\longleftarrow}}
  ~\equiv~
 {{\stackrel{q}\longrightarrow}\atop{\stackrel{q+l}\longleftarrow}}
 \bareDA\glue\bareDA
 {{\stackrel{p}\longrightarrow}\atop{\stackrel{p+l}\longleftarrow}}
\ .
\ee
We want to argue that due to the transversal projector 
in $\DA$, this and all similar diagrams are finite. Up to an 
overall factor, and integrating over the time difference between
the two vertices, they are
\bea \label{D correction}
\int\frac{\rmd^d k}{(2\pi)^d} \, 
\left[\frac1{(k^2+m^2)^{\frac{d+\E}{2}}}\right]^{2}\,
\frac{[q(q+l)][(k+q)^2] - [q(k+q)][(q+l)(k+q)] }{(k+q)^2} \nn\\
\times\frac{[p(p+l)][(k+p)^2] - [p(k+p)][(p+l)(k+p)] }{(k+p)^2}
\ .
\eea
Since  for large $k$
\bea
\frac{[q(q+l)][(k+q)^2] - [q(k+q)][(q+l)(k+q)] }{(k+q)^2} &=& O(k^0) \nn\\
\frac{[p(p+l)][(k+p)^2] - [p(k+p)][(p+l)(k+p)] }{(k+p)^2} &=& O(k^0)
\eea
the integral \eq{D correction}
scales for large $k$ as 
\be
\int\frac{\rmd^d k}{(2\pi)^d}\,
\left[\frac1{(k^2+m^2)^{\frac{d+\E}{2}}}\right]^{2} \sim \int \frac {\rmd k}k 
\frac 1{k^{d+2\E}}
\ee
and is thus $UV$-convergent for any $d$ and $\E>0$. This means that
$Z_D=1$. Note that the 
transversal projectors ensure that no additional divergences appear for
$\E=0$ at $d=2$ or $4$, since it allows to bring the derivatives always
to the external legs. Moreover, no long-range interaction can be generated.
This argument can be generalized to higher orders in perturbation theory,
however only the absence of additional divergences for $d>2$ is immediately
apparent. We shall not elaborate on this question any longer, since it
is not at the center of our analysis.

Let us now come back to counter-terms for $\nu$. By direct inspection, 
one sees that the only diverging diagrams are chains of bubbles, of the form
\be
\barePA\ ,\quad \barePA\glue\barePA  ,\quad \barePA\glue\barePA\glue\barePA\ , \ldots
\ .
\ee
However, these diagrams are already renormalized by \Eq{s:Znu}.
This is easily seen by directly summing the perturbative (geometric)
series, as e.g. in \cite{Wiese1998a}.
The $\beta$-function  to all orders in perturbation theory thus reads
\be \label{s:beta}
\beta(u) := \AT{m\frac\p{\p m} }{0} u = -\E u + \left(1-\frac1d\right)
\frac{C_d}2 u^2
\ .
\ee
(Note that we do not take $C_d$ at $\E=0$; this ``minimal subtraction''
is completely sufficient at 1-loop order, but unsufficient for the 
all order result \eq{s:beta}.)
This $\beta$-function has a fixed point at 
\be \label{s:u star}
	u^*=\frac{2d}{(d-1)C_d}\, \E\ .
\ee
One can now define the anomalous dimension $\gamma_\nu$ of $\nu$
as
\bea
\gamma_\nu(u)&:=& m \frac \p {\p m} \ln Z_\nu \nn\\
&=&u \left(1-\frac1d\right) \frac {C_d}2
\label{s:gamma def}
\ ,
\eea
which at $u=u^*$ reads (to all orders in $\E$)
\be \label{s:gamma star}
\gamma_\nu^*=\E \ .
\ee
Since $G_f$ is not renormalized, this leads
 to an anomalous dimension of $\Theta$ (in units of $x\sim 1/m$)
to all orders in perturbation theory as
\be \label{s:eta star}
\eta^*=-\frac\E2 \ .
\ee
The full dimension of $\Theta$ thus is 
\be
\left[ \Theta \right]_{x,\ind f} = 1-\frac\E2\ .
\ee 
This simple scaling is only part of the whole story, 
as we shall see in the next section. 

To summarize: we have constructed a renormalized action which is 
UV-finite in the limit of $\E\to0$, and which gives the IR-scaling
for  $\E>0$.  

\subsection{Observables and IR-divergences}
We now want to study correlation-functions as e.g.
\be \label{s:S2n def}
S^{2n}(x-y,t):=\langle [\Theta(x,t)-\Theta(y,t)]^{2n}\rangle\ .
\ee
We always choose $x$ and $y$ inside the injection region, 
thus especially $L\gg |x-y|$.
It will turn out that these observables are sensitive to the 
size $L=1/M$ of the system, and demand new renormalizations. 
Since from the viewpoint of $\phi^4$-theory this is rather strange, 
let us study an expectation value in the latter theory
 in order to see where the
difference to the passive scalar problem 
lies. Suppose, one wants to evaluate the 
expectation value 
\be \label{s:U def}
U(x,y):=\half \left< \phi^2(x) \phi^2(y) \right> \ ,
\ee
for the theory defined by the Hamiltonian in $d$ dimensions
\be
{\cal H} [\phi] = \int\rmd^d x\, \half (\nabla \phi(x))^2+ b  \phi^4(x)
\ .
\ee
(For the difference in definition between  \Eqs{s:S2n def} and 
\eq{s:U def} note that for correlation functions growing with the 
distance, definition \eq{s:S2n def} has to be used, whereas for 
decaying correlation functions, \eq{s:S2n def} is the correct one.)

Denoting expectation values in the free theory by 
$C(x-y):=\left<\phi(x) \phi(y) \right>_0\sim |x-y|^{2-d}$, 
the first contributions to $U(x,y)$ are (when setting
to 0 self-contractions in the  $\phi^4$-interaction, and neglecting 
combinatorial factors)
\be
U(x,y) = \PhiD 
- b \PhiA +b^2 \!\left(\!\PhiB +\PhiC\!\right) + O(b^3) \ .
\ee
This formulas is to be understood such that the outer points are 
always $x$ and $y$ and that one integrates over the inner points. 
Since $C(s)\sim s^{2-d}$ the term of order $b$ scales as
\be
\PhiA \sim \int\rmd^dz\, |x-z|^{4-2d}\,|y-z|^{4-2d} \ ,
\ee
which for  for large $z$ becomes
\be \label{s:conv}
\int\rmd^dz\,  |z|^{8-4d} \ .
\ee
It is IR-convergent at least for $d$ close to $4$. 
Now still consider one of the terms of order $b^2$. 
\be \label{2.36}
\PhiB = \int\rmd^ds\, \rmd^dt\, C(x-s)^2 C(s-t)^2 C(t-y)^2
\ .
\ee
Similar to what has happened  in the last section, there is 
a logarithmic divergence at $\E=0$ for small $s-t$, 
which has to be renormalized. Calculating directly the integral over
$s-t$ in the regularized theory at $d<4$ , this leads to ($z:=s-t$) 
\be \label{s:2.32}
_s\PhiD_t = \int\rmd^d z \, C(z)^2 \sim \frac 1{4-d} \, L^{4-d}\ ,
\ee
where $L$ is an effective IR-cutoff. The question now arises, what
$L$ is. Noting that the integral over the center of mass $(s+t)/2$
is IR-convergent with the identical argument that led to \Eq{s:conv},
the effective scale at which the integral \eq{s:2.32} is cut off, 
is $L=|x-y|$. These kind of arguments can be continued to 
higher orders\footnote{A caveat is in order here: Calculating at
small but finite
values of $\epsilon:=4-d$, there is always an IR-divergence at sufficiently
high orders in perturbation theory. To see this take a long chain of 
bubbles, similar to that of \Eq{2.36}. The standard way to 
circumvent this well-known 
(technical) problem of the massless theory is to
take  $\E$ small enough inside each diagram.}. 
They show three things: First, expectation values of 
physical observables are IR-finite, i.e.\ boundaries of 
the system do not enter into their calculation.  Second, the distances between 
the observable points set the largest scale $L$ in the problem.
Third, when varying these distances, $L$ changes and thus the 
value of diverging subdiagrams as \eq{s:2.32}. This  gives rise 
to an anomalous scaling of the observables. The latter is most
comfortably taken care of by the renormalization group procedure, 
which also allows for  a proof of the above statements. 

\smallskip
Let us now turn back to the passive scalar problem, and consider 
\be
S^2(x-y,t=0) := \left< [\Theta(x,0)-\Theta(y,0)]^2 \right>\ .
\ee
The order 0 contribution is
\bea
\PC = \int_0^{\infty}\rmd t\, \int\rmd^dz\, \rmd^dz'&\left[R(x-z,t)-
R(y-z,t)\right]~& \nn\\
\times& \left[R(x-z',t)-
R(y-z',t)\right]& G_f(M|z-z'|)\qquad\ \ 
\ .
\eea
Using \Eq{s:R(x,t)} this can be written as 
\bea \label{s:i1}
\lefteqn{\PC } \nn\\
	&&{\sim \int\rmd^dz\, \rmd^dz'
	\bigg( \left[ (x{-}z)^2 {+}(x{-}z')^2\right]^{1{-}d}}{+}
	\left[ (y{-}z)^2 {+}(y{-}z')^2\right]^{1{-}d} 
	{-}\left[ (x{-}z)^2 {+}(y{-}z')^2\right]^{1{-}d} \nn\\
	&&\hphantom{\sim \int\rmd^dz\, \rmd^dz'
	\bigg( \left[ (x{-}z)^2 {+}(x{-}z')^2\right]^{1{-}d}}
	{-}\left[ (y{-}z)^2 {+}(x{-}z')^2\right]^{1{-}d}
	\bigg)
	 G_f(M|z{-}z'|) \ .	
\eea
Integrating over both $z$ and $z'$ large, the integral \eq{s:i1} scales as 
\be \label{s: log div}
\int\rmd^{2d} s\, s^{2(1-d)} s^{-2} G_f(Ms) \sim\
\int\frac{\rmd s}s\,  G_f(s) \sim \ln L \ ,
\ee
where the factor of $s^{-2}$ is due to the 
differences in \Eq{s:i1}, and $L=1/M$ is the scale at which $G_f$ starts
to decay rapidly. $S^2(x-y,t)$ thus {\em explicitly} depends on the largest 
distance $L>|x-y|$ in the problem, in the very contrast to 
the example of the $\phi^4$-theory considered above. 

We shall now  show that perturbative corrections in $\rD$ make 
$S^2$ depend even stronger on $L$, namely
as  $L^{n\E}$ at $n$th order in perturbation theory. 
To this aim consider the term of order $\rD$. Using the regularized theory,
it appears at two places: First $\nu$ is renormalized, and thus the 
resulting effective response-function $R(x,t)$ 
decays faster -- this effect would render \eq{s: log div} IR-convergent.
However, there is a second contribution, namely
\bea \label{s:PE}
\rD \PE=\int\rmd^d z \int \rmd^d z'\int \rmd t\, 
\frac{\partial}{\p z_i}\left[ R(x-z,t)-R(y-z,t) \right]&  \nn\\
\times \frac{\partial}{\p z'_j}\left[ R(x-z',t)-R(y-z',t) \right]&
\rD_v^{ij}(x-y)
\ .
\eea
This is only some part of the diagram, and in principle, 
it has to be closed through a $\PD$, leading to 
\be \label{s:2.43}
\PF \ .
\ee
Since this diagram is again plagued by an IR-divergence, 
the leading contribution in \Eq{s:PE} will come from the domain 
of large $z$ and $z'$. In that limit, $|x-y|$ is much smaller
than both $z$ and $z'$ and consequently, 
$\Theta(x)-\Theta(y)$ can be replaced by $(x-y)\nabla \Theta(\frac{x+y}2)$.
Let us again stress that this is valid in the domain, where 
$|x|\ll L$ and $|y| \ll L$, or when using $G_f(M|x-y|)$ where
$|x-y|\ll L$.
Due to that replacement, we can now use a very powerful trick:
Instead of analyzing the IR-divergences of $\langle [\Theta(x,t)-\Theta(y,t)]^2
\rangle$, or more generally of 
$S^{2n}(x-y,t):=\langle [\Theta(x,t)-\Theta(y,t)]^{2n}
\rangle$
we can analyse the {\em UV-divergences} of
 the {\em composite
operator} $[(x-y)\nabla \Theta(\frac{x+y}2)]^{2n}$. 
The latter however is a standard task in perturbative renormalization. 
We will see in the next section, that this leads to a whole family
of operators and anomalous dimensions; the operator with the smallest
dimension will then give the term which most sensitively depends on 
$L$.  In order to avoid confusions, let us already note
that the second moment $S^{2}(r)$ discussed above does not depend
on $L$, since the contribution to the response-function and that 
in \Eq{s:2.43} just cancel. This can also be obtained exactly
\cite{Kraichnan1968}.

\subsection{The scaling of $S^{2n}$ and renormalization of
composite operators}\label{s:scaling of S 2n}
As discussed in the last subsection, we
 now have  to study the renormalization of 
 $[z\nabla \Theta]^{2n}$. It will turn out that under renormalization
this term generates $[z\nabla \Theta]^{2n-2} z^2 [(\nabla \Theta)^2]$.
In a second step, $[z\nabla \Theta]^{2n-2} z^2 [(\nabla \Theta)^2]$
generates $[z\nabla \Theta]^{2n-4} z^4 [(\nabla \Theta)^2]^2$ a.s.o.\ until also a term of the form $ z^{2n} [(\nabla \Theta)^2]^n$ is 
generated. All these operators will mix under renormalization. The eigen operator
with the smallest dimension will give the term which most strongly depends
on $L$. 

We now treat the general case. Define
\be
{\cal S}^{(n,m)}:=  z^{2n}\, [(\nabla \Theta)^2]^n\, [z\nabla \Theta]^{2m}
\ .
\ee
We first observe that the operator product expansion (denoted by 
$\contract$) is
\be \label{s:2.41}
{\cal S}^{(n,m)} \contract \frac \rD2 \DA =
T^{ij} \left[ (\nabla^i\Theta) (\nabla^j\Theta) \contract \frac \rD2 \DA 
\right]
\ee
with
\bea
T^{ij}&=& \half \frac{\p}{\p(\nabla^i\Theta)}\frac{\p}{\p(\nabla^j\Theta)}
\left\{z^{2n} \left[ (\nabla \Theta)^2 \right]^{n}  (z\nabla \Theta)^{2m}
 \right\}\nn\\
&=&n  \delta^{ij} z^{2n}\left[ (\nabla \Theta)^2\right]^{n-1}
(z\nabla \Theta)^{2m}
+ 2n(n-1)z^{2n}(\nabla^i\Theta)(\nabla^j\Theta)
\left[ (\nabla \Theta)^2 \right]^{n-2}  (z\nabla \Theta)^{2m}
\nn\\
&&+2nm z^{2n}\left[ z^i \nabla^j\Theta + z^j \nabla^i\Theta \right]
\left[ (\nabla \Theta)^2 \right]^{n-1}  (z\nabla \Theta)^{2m-1}
\nn\\
&&+ m (2m-1) z^{2n}
z^iz^j
\left[ (\nabla \Theta)^2 \right]^{n}  (z\nabla \Theta)^{2m-2} \ .
\label{s:Tij}
\eea
\Eq{s:2.41} then reads
\be
\rD \int_0^\infty \rmd t \int\! \frac{\rmd^d p}{(2\pi)^d}\, T^{ij}p^ip^j R^2(p,t) 
\frac{1}{(p^2+m^2)^{\frac{d+\E}{2}}}
\left[ (\nabla \Theta)^2 - \frac{(p \nabla \Theta)^2}{p^2} \right]
+ O(\E^0) \ .
\ee
Since $R(p,t)=\rme^{-\nu_0 p^2 t}\Theta(t)$, integration over $t$
yields (up to finite terms)
\be \label{s:last}
\frac{\rD}{2\nu_0}  \int\! \frac{\rmd^d p}{(2\pi)^d}\, T^{ij} \frac{p^ip^j}{p^2} 
\frac{1}{(p^2+m^2)^{\frac{d+\E}{2}}}
\left[ (\nabla \Theta)^2 - \frac{(p \nabla \Theta)^2}{p^2} \right]
\ee
$T^{ij}$ in \Eq{s:Tij} has the form 
\be \label{TijABC}
T^{ij}= A^2 \delta^{ij}+B^i C^j\ .
\ee
Inserting this into \Eq{s:last} and using the formulas from 
appendix \ref{a:Some integrals} and \Eq{s:Cd def}	yields
\be
\frac u 2 {C_d} \frac{m^{-\E}}{\E} 
\left[ 
A^2 (\nabla \Theta)^2 \left(1{-}\frac 1 d\right)\! + (\nabla \Theta)^2
(BC)\frac1d -\frac1{d(d{+}2)}\left( 2 (B\nabla\Theta) (C\nabla\Theta)
{+}(BC) (\nabla \Theta)^2\right)
\right]
\ .
\ee
Specifying $T^{ij}$ in \Eq{TijABC} to its value of \Eq{s:Tij} gives
\bea
\frac{u}2 {C_d} \frac{m^{-\E}} \E \frac{1}{d(d+2)}
\bigg\{
&&\!\!\left[n(d-1)(d+2n+4m)-2m(2m-1)\right]  
z^{2n}\, [(\nabla \Theta)^2]^n\, [z\nabla \Theta]^{2m}
\nn\\
&&+ m(2m-1)(d+1) z^{2n+2}\, [(\nabla \Theta)^2]^{n+1}\, [z\nabla \Theta]^{2m-2}\bigg\}\ .
\eea
The final result when 
contracting ${\cal S}^{(n,m)}$ with $\frac \rD2 \DA$ is
\bea \label{s:final MOPE}
{\cal S}^{(n,m)} \contract \frac \rD2 \DA &=&
{\cal S}^{(n,m)} \frac{u}2 \frac{C_d}\E \,m^{-\E}\frac{1}{d(d+2)}
\left[n(d-1)(d+2n+4m)-2m(2m-1)\right] \nn\\
&&+{\cal S}^{(n+1,m-1)}\frac{u}2 \frac{C_d}\E \,m^{-\E}\frac{d+1}{d(d+2)}\,
m(2m-1) \ .
\eea
This shows that ${\cal S}^{(0,n)}, {\cal S}^{(1,n-1)}, \ldots,
{\cal S}^{(n,0)}$ mix under renormalization. 
The first two eigen operators are ${\cal S}^{(n,0)}$ and 
$\bar{\cal S}^{(n-1,1)}:=
{\cal S}^{(n-1,1)}-\frac{1}{d}{\cal S}^{(n,0)}$, 
with eigen values
\bea
{\cal S}^{(n,0)} \contract \frac \rD2 \DA &=&
{\cal S}^{(n,0)} \frac{u}2 \frac{C_d}\E \,m^{-\E}\frac{n(d-1)(d+2n)}{d(d+2)}
\\
\bar{\cal S}^{(n-1,1)} \contract \frac \rD2 \DA &=&
\bar{\cal S}^{(n-1,1)} \frac{u}2 \frac{C_d}\E \,m^{-\E}\frac{(n-1)(d-1)(d+2n+2)-2}{d(d+2)} 
\ .
\eea
More importantly, due to the triangular form of the matrix, 
the eigen values can just be read off from the diagonal. 
Therefore the eigen operators $\bar{\cal S}^{(n,m)}$ are multiplicatively
renormalizable (again denoting by index $_0$ bare quantities)
\bea
\bar{\cal S}_0^{(n,m)} &=& Z^{(n,m)} \bar{\cal S}^{(n,m)} \nn\\
Z^{(n,m)} &=& 1- \frac{u}2 \frac{C_d}\E \frac{1}{d(d+2)}
\left[n(d-1)(d+2n+4m)-2m(2m-1)\right] 
\label{s:2.55}
\ .
\eea
This yields the anomalous scaling-function $\gamma^{(n,m)}(u)$ 
of $\bar{\cal S}_0^{(n,m)}$
in units of $x$ as 
\bea \label{s:gamma(n,m)(u)}
\gamma^{(n,m)}(u) &:=& -m \frac\p{\p m} \ln Z^{(n,m)} \nn\\
	&=&	-\frac u 2 C_d  \frac{1}{d(d+2)}
\left[n(d-1)(d+2n+4m)-2m(2m-1)\right]
\ .
\eea
At the IR fixed point $u=u^*$ from \Eq{s:u star}, this is
\be \label{s:gamma(n,m)}
\gamma^{(n,m)} =	-  \frac{\E}{(d-1)(d+2)}
\left[n(d-1)(d+2n+4m)-2m(2m-1)\right]
\ .
\ee
Taking care of the naive perturbative contribution to 
the scaling of $\bar{\cal S}_0^{(n,m)}$ from \Eq{s:eta star},  
we finally obtain the full scaling-dimension $\Delta^{(n,m)}$ of 
$\bar S_0^{(n,m)}$
\bea 
\Delta^{(n,m)} &=&  \gamma^{(n,m)}-2(n+m)\eta^*\nn\\
&=&\gamma^{(n,m)} +(n+m) \E \nn\\
&=& -\E \frac{2n(n-1+2m)}{(d+2)} + \E \frac {m(4m-1)+d^2+d}{(d+2)(d-1)}
\label{s:Delta(n,m)}
\ .
\eea
These results have already been obtained in 
\cite{GawedzkiKupiainen1995,BernardGawedzkiKupiainen1996,%
AdzhemyanAntonovVasilev1998}, where only the case $m=0$ was given.
The case $m>0$ can be found in \cite{Antonov1998}.

The exponents satisfy the inequality
\be
\Delta^{(n,m)}<\Delta^{(n-1,m+1)} \ ,
\ee
such that indeed ${\Delta^{n}}:=\Delta^{(n,0)}$ is 
the smallest of all exponents
$\Delta^{(n-m,m)}$ and dominates the $L$-dependence of $S^{2n}(r,t)$
as stated above. 
Explicitly, 
\be \label{s:explicit}
S^{2n}(r,t) ~\sim~ r^{n(2-\E)} \BRA{\frac{r}{L}}{\Delta^{n}} \ ,
\qquad \Delta^{n}=-\E \frac{2n(n-1)}{d+2} \ .
\ee 
Only the second moment does not depend on $L$. As demonstrated in 
\cite{Kraichnan1968}, this is the consequence of a conservation
law, which allows for an exact calculation of the second moment. 

These results have been tested numerically with different 
methods in \cite{KraichnanYakhotChen1995,FairhallGalantiLvovProcaccia1997,%
ChenKraichnan1998,FrischMazzinoNoullezVergassola1999}.

The other question one might ask is why only one of the two factors
in \Eq{s:explicit} depends on $L$, and whether the $r$-dependence
comes out correctly. 
To understand this point, we recall that the first factor is due to 
the renormalization of $\nu$, and thus contributes to the anomalous 
dimension of $\Theta$, irrespective of the boundary conditions.
 The second factor stems from the anomalous
dimension of the composite operator
$S^{(n,0)}$, which was associated to the IR-divergence, i.e.\ 
$L$-dependence of $S^{2n}(r,t)$, and which has two contributions: the proper
renormalization of $S^{(n,0)}$ as given by $Z^{(n,0)}$ or $\gamma^{(n,0)}$,
and the renormalization of $\nu$; these add up to $\Delta^{(n,0)}$
as given in \Eq{s:Delta(n,m)}.
Only the combination of these terms
 contribute to the $L$-dependence of $S^{2n}(r,t)$.

Also note that the exponents with $m>0$ are also observable, and 
correspond to observables of different symmetries.

\maybeinput{langevin.tex}{Langevin-description of the passive scalar}			
\maybeinput{membrane.tex}{Generalization to polymers and membranes}			
\maybeinput{conclusion.tex}{Conclusions}		
\appendix 
\maybeinput{appendices.tex}{Appendices}		
\section*{Acknowledgements}
It is a pleasure to thank Terry Hwa and Pierre Le Doussal for introducing
me to the problem. I have much learned from discussions with
both of them as well as with Denis Bernard, Jens Eggers, 
Stefan Thomae and Massimo Vergassola.
 Partial financial support from UCSD is acknowledged.

\protect\markboth{K.\,J.~Wiese, The passive polymer problem}{References}
  
\setcounter{section}{17} 
\section{References}
\def\refname{} \vspace*{-1.2cm}
\bibliography{../../citation/citation}
\bibliographystyle{../../macros/KAY}


\end{document}